\documentclass[letter,showpacs,prb,final,groupedaddres,twocolumn]{revtex4-1}
\usepackage{graphicx}
\usepackage{amssymb}
\usepackage{marvosym}
\usepackage{latexsym}
\usepackage[usenames]{color}
\usepackage{float}
\usepackage{array}
\usepackage{verbatim} 
\usepackage{natbib} 
\usepackage{hyperref}

\begin{document}

\title{Collapsed tetragonal phase transition in LaRu$_2$P$_2$}

\author{Gil Drachuck}

\author{Aashish Sapkota}

\author{Wageesha T. Jayasekara}

\author{Karunakar Kothapalli}

\author{Sergey L. Bud'ko}

\author{Alan I. Goldman}

\author{Andreas Kreyssig}

\author{Paul C. Canfield}

\affiliation{Department of Physics and Astronomy and Ames Laboratory, Iowa State University, Ames, IA 50011, USA}

\date{\today}

\begin{abstract}

The structural properties of LaRu$_2$P$_2$ under external pressure have been studied up to 14 GPa, employing high-energy x-ray diffraction in a diamond-anvil pressure cell. At ambient conditions, LaRu$_2$P$_2$ (\textit{I}4/\textit{mmm}) has a tetragonal structure with a bulk modulus of  $B=105(2)$~GPa\textbf{} and exhibits superconductivity at $T_c= 4.1$~K.  With the application of pressure, LaRu$_2$P$_2$ undergoes a phase transition to a collapsed tetragonal (cT) state  with a bulk modulus of $B=175(5)$~GPa. At the transition, the \textit{c}-lattice parameter exhibits a sharp decrease with a concurrent increase of the \textit{a}-lattice parameter. The cT phase transition in LaRu$_2$P$_2$ is consistent with a second order transition, and was found to be temperature dependent, increasing from $P=3.9(3)$~GPa at 160~K to $P=4.6(3)$~GPa at 300~K. In total, our data are consistent with the cT transition being near, but slightly above 2~GPa at 5~K where superconductivity is suppressed. Finally, we compare the effect of physical and chemical pressure in the \textit{R}Ru$_2$P$_2$~(\textit{R}~=~Y,~La$\textendash$Er,~Yb) isostructural series of compounds and find them to be analogous. 
\end{abstract}
%\pacs{74.62.Bf, 81.40.Vw }

\maketitle

\section{Introduction}
%\label{Introduction}
 
Compounds with the ThCr$_2$Si$_2$ structure have generated a lot of interest over the years, most recently, the discovery of high-temperature superconductivity (SC) in the \textit{Ae}Fe$_2$As$_2$ (\textit{Ae}=Ba, Sr, Ca) family.\cite{Canfield2010,Johnston2010} The physical properties in this family of superconductors can be tuned typically via chemical substitution, or by external pressure, from a magnetic/orthorhombic phase to SC.\cite{Rotter2008,Alireza2009,Nini2009,Paglione2010,Stewart2011,Canfield2010} At large enough applied pressures, a collapsed tetragonal (cT) phase transition occurs, in which the \textit{c}-lattice parameter can decreases by up to 10\% in CaFe$_2$As$_2$.\cite{Kreyssig2008,Canfield2009} The critical pressure for this transition ranges from 0.4~GPa in CaFe$_2$As$_2$~\cite{Kreyssig2008,Yu2009} to 10~GPa and 17~GPa for SrFe$_2$As$_2$~\cite{Kasinathan2011} and BaFe$_2$As$_2$~\cite{Uhoya2010}, respectively. However, the cT phase transition is not unique to the \textit{Ae}Fe$_2$As$_2$ family and was predicted in compounds with the \textit{A}\textit{B}$_2$\textit{X}$_2$ (\textit{B} = transition metal, \textit{X} = group
14 or 15 element) structure in which the \textit{X}-\textit{X} contact varies over the range of bonding between no bond and a fully formed \textit{X}-\textit{X} single bond. \cite{Hoffmann1985} Indeed, the cT phase has been observed in many compounds with the ThCr$_2$Si$_2$ structure~\cite{Huhnt1998,Bishop2010,Yu2014,Naumov2017} as well as in the recently discovered CaKFe$_4$As$_4$ superconductor which may even host a two-step cT transition.\cite{Kaluarachchi2017}

The electronic properties of the cT phase are vastly different than of the un-collapsed tetragonal (TET) phase due to the abrupt change of the electronic band-structure, associated with the structural change~\cite{Dhaka2014}, e.g., the transition into the cT phase often leads to the loss of magnetism or SC.\cite{Kreyssig2008,Canfield2009,Ran2012,Gati2012,Furukawa2014,Zhang2016} Therefore, studying the pressure-temperature phase diagrams and the different ground states is fundamental for understanding SC this family of materials. Some compounds with ThCr$_2$Si$_2$ structure can support conventional superconductivity while still exhibiting intriguing ground states. One example is the compound LaRu$_2$P$_2$, in which it had been shown that SC in LaRu$_2$P$_2$ is enhanced under external pressure, but then either vanishes~\cite{Foroozani2014} or diminishes for pressures greater than $P=2.1$~GPa, at which a cT transition has been predicted by band-structure calculations.\cite{Li2016}
  
\begin{figure*}[t]
 	\begin{center}
 		\includegraphics[width=172mm, trim=0cm 0cm 0cm 0cm]{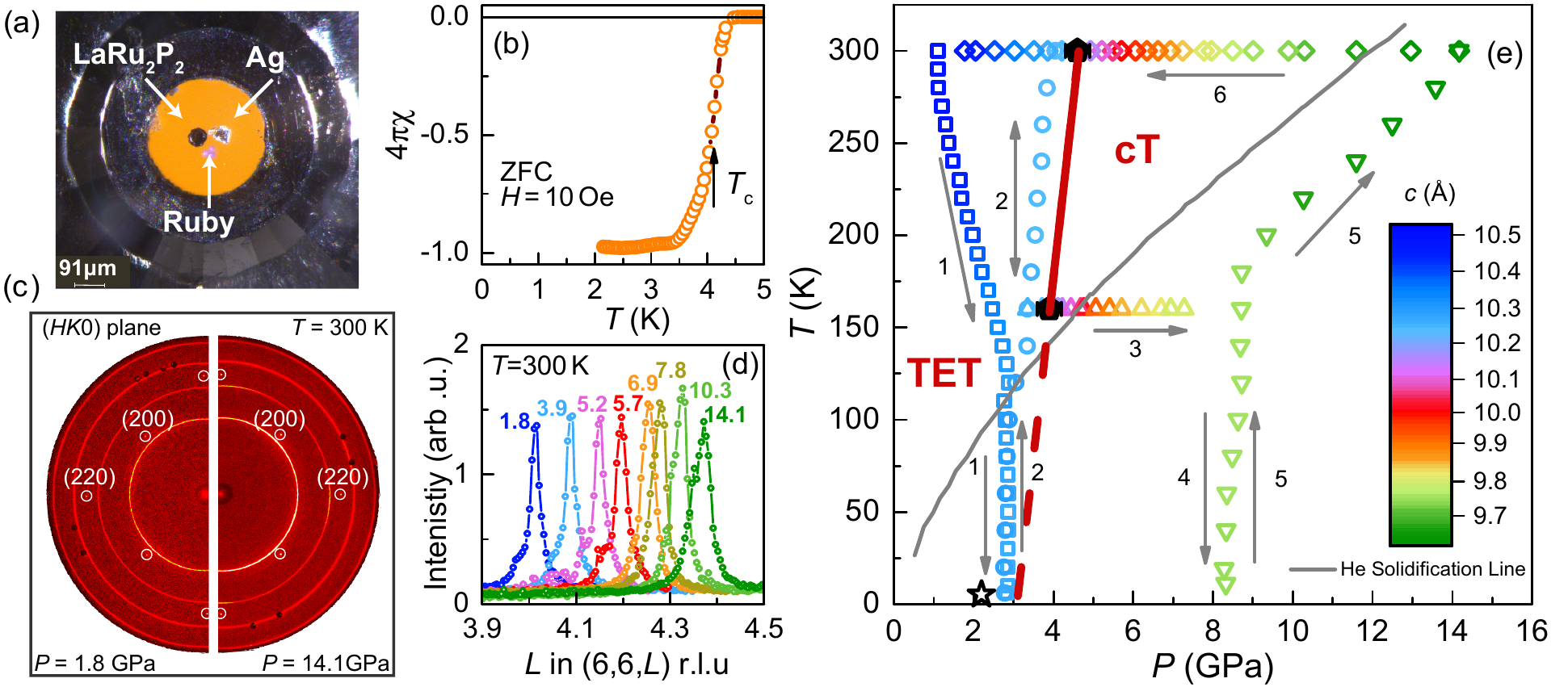}
 	\end{center}
 	\caption{(a) Image of the LaRu$_2$P$_2$ single crystal on which x-ray diffraction data were measured, ruby spheres and a piece of silver which were used separately as in-situ pressure gauges, mounted in a diamond anvil cell, before helium gas loading. (b) DC magnetization measured on several LaRu$_2$P$_2$ crystals showing the superconducting transition at $T_c=4.1(2)$~K as indicated by the arrow. (c) Diffraction pattern of the (\textit{HK}0) plane acquired at $T=300$~K, by the MAR345 image plate detector for $P=1.8$~GPa (left) and $P=14.1$~GPa (right). The (\textit{HK}0) Bragg peaks are labeled and marked by white circles. The silver Bragg rings were used to determine the pressure during the experiment. (d) Several raw data curves of the (664) Bragg peaks, from which the \textit{c}-lattice parameters were determined for different pressures (in GPa). (\textit{HKL}) values are given relative to the lattice parameter at 300~K and 1.8~GPa. (e) The temperature-pressure points at which structural data on LaRu$_2$P$_2$ were collected. The open symbols represent $\left(T,P\right)$ points at which lattice parameters were measured, color-coded with the \textit{c}-lattice parameter value to emphasize the pressure induced collapsed tetragonal transition in LaRu$_2$P$_2$. The closed black symbols represent the collapsed tetragonal transition pressures for two isothermal pressure scans. The red line represents the TET/cT phase boundary. The previously reported suppression of superconductivity in LaRu$_2$P$_2$ at $P=2.1$~GPa~\cite{Foroozani2014} is indicated by the open black star, clearly in proximity to the TET/cT transition. The gray line indicates the He solidification line.}
 	\label{Fig1}
 \end{figure*}

LaRu$_2$P$_2$ is part of the \textit{R}Ru$_2$P$_2$ (\textit{R} = Y, La$\textendash$Er, Yb) series of compounds which crystallize in the ThCr$_2$Si$_2$ type structure (space group \textit{I}4/\textit{mmm}),  with lattice parameters $a=4.031 \AA$ and $c=10.675 \AA$. ~\cite{Jeitschko1987} It goes through a superconducting transition at $T_c= 4.1$~K [Fig.~\ref{Fig1}(b)]. LaRu$_2$P$_2$ is isostructural to the \textit{Ae}Fe$_2$As$_2$, however Ru is not moment bearing in this compound. SC in LaRu$_2$P$_2$ has a different origin with respect to the \textit{Ae}Fe$_2$As$_2$ family of superconductors.\cite{RazzoliPRL2012} In this compound, the superconducting properties are isotropic, with higher carrier density than in the iron-based superconductors.\cite{Ying2010} Band-structure calculations show that the electronic properties of LaRu$_2$P$_2$ exhibit three-dimensional rather than two-dimensional characteristics found in the \textit{Ae}Fe$_2$As$_2$ family, and that $T_c$ can be well estimated from the size of electron-phonon coupling using BCS theory.\cite{Karaca2016} The effect of hydrostatic pressures on the SC properties of LaRu$_2$P$_2$ were previously explored by extensive magnetization~\cite{Foroozani2014} and electrical transport measurements together with band-structure calculations~\cite{Li2016}. They find an initial increase in the  $T_c$ up to $P=2.1$~GPa followed by an abrupt disappearing of the Meissner response accompanied by a broadening of the transition in resistivity.   

In this work, we applied hydrostatic pressure up to 14~GPa, using a diamond-anvil cell, on a single crystal of LaRu$_2$P$_2$ and studied its structural properties using high-energy x-ray diffraction. This allowed us to address the reports showing a change of superconducting properties around $P=2.1$~GPa~\cite{Foroozani2014,Li2016} and to compare the effect of hydrostatic (physical) to chemical pressure upon substitution of different rare-earth elements, \textit{R}, in \textit{R}Ru$_2$P$_2$, associated with the reduction of ionic radii, \textit{i.e.}, lanthanide contraction.\cite{Lanthanide}

\section{Experimental}

Single crystals of LaRu$_2$P$_2$ were grown out of a high-temperature solution rich in Sn. Stoichiometric quantities of pieces of elemental La, Ru and P were mixed with Sn in molar ratios between 1:10 to 1:80 (LaRu$_2$P$_2$ to Sn).\cite{CanfieldFisher2001,CanfieldGrowth} Different dilution levels were attempted in order to optimize the planar size of grown crystals for the diamond-anvil pressure cell requirements ($<60\mu$m). Small crystals of that approximate size were abundant in the most dilute growth (1:80). The initial elements were placed into the bottom 2~ml alumina crucible of a Canfield Crucible Set\cite{CanfieldFrit}, and sealed in amorphous silica ampules under a partial argon atmosphere. The ampules were heated to 300~$^\circ$C in 3 hours and dwelled there for 6 hours, in order to allow the phosphorous and tin to react, therefore reducing the risk of explosions upon further heating. Subsequently, the ampules were heated over 10 hours to 1190~$^\circ$C where they dwelled for 3 additional hours, then cooled, over 250 hours to 780~$^\circ$C. At that point, the excess molten Sn-rich solution was decanted by a modified centrifuge.~\cite{CanfieldGrowth,CanfieldFrit} The grown crystals had square plate-like morphology with the \textit{c}-axis perpendicular to the plate surface with dimensions ranging from 40 to 200~$\mu$m. 

To confirm that the grown crystals are in fact superconducting, DC magnetization was measured in a Quantum Design Magnetic Property Measurement System (MPMS), SQUID magnetometer, on several LaRu$_2$P$_2$ crystals. In Fig.~\ref{Fig1}(b) we show zero-field cooled magnetization data measured  under an applied magnetic field of  $H=10$~Oe. The magnetization exhibits a superconducting transition at $T_c= 4.1(2
)$~K (mid-point of the transition), which is consistent with previously reported values.\cite{Jeitschko1987,Foroozani2014,Li2016} %the small shoulder is apparent at $T=3.2$~K which is associated to the superconducting transition of the residual tin flux. 

High-energy x-ray diffraction measurements were performed on the six-circle diffractometer at the 6-ID-D station at the Advanced Photon Source, using 100.32~keV x-rays and a beam size of $100 \times 100$ ~$\mu$m$^2$. A double-membrane-driven~\cite{Sinogeikin2015} copper-beryllium diamond-anvil cell (DAC) with 600~$\mu$m culet anvils was used to generate high pressures. A steel gasket was pre-indented to 65~$\mu$m thickness and a hole of diameter 260~$\mu$m was laser-drilled~\cite{doi:10.1063/1.4926889} to serve as sample chamber. A single crystal with dimensions of $52\times52\times22$~$\mu$m$^3$ was placed in the sample chamber together with ruby spheres and silver foil for pressure calibration as shown in Fig.~\ref{Fig1}(a). Helium gas, loaded at $P=1.1$~GPa, was used as the pressure-transmitting medium. The pressure was initially determined from fluorescence lines of the ruby spheres at ambient temperature. During the diffraction measurements, the pressure was determined \textit{in-situ} by analyzing selected Bragg peaks from the silver foil. The DAC was mounted on the cold finger of a He closed-cycle cryostat and temperature-dependent measurements were performed between $T=5$ and 300~K for various pressures.

The x-ray diffraction patterns of the (\textit{H}\textit{K}0) plane, from which the \textit{a}-axis parameter was determined, were recorded using a MAR345 image plate detector positioned at 1.486~m behind the sample, as the DAC was rocked by up to $\pm 3.6^{\circ}$ about two independent axes perpendicular to the incident x-ray beam.~\cite{Kreyssig2007} High-resolution diffraction patterns of the (66\textit{L}) Bragg reflections, from which the \textit{c}-axis parameter was obtained, were recorded by  a Pixirad-1 detector positioned 1.397~m behind the sample, at the appropriate diffraction angle, while rocking around one of the two axes perpendicular to the x-ray beam. The lattice parameters were determined by fitting the Bragg peak positions after integrating the data over the transverse scattering directions. This procedure was used for both the data recorded by the Pixirad-1 detector and the MAR345 image plate system.

%DC magnetization measurements were performed in a Quantum Design Magnetic Property Measurement System (MPMS), superconducting quantum interference device (SQUID) magnetometer (\textit{T} = 1.8 - 300 K, \textit{H}$_{\mathrm{max}}$ = 55 kOe). The crystals were place into a gel capsule, whose background is negligible compared to the superconducting diamagnetic signal at low magnetic fields.  

\section{Results}
%\label{Results}

In Fig.~\ref{Fig1}(c) we present raw x-ray diffraction patterns of the (\textit{HK}0) plane, acquired at $T=300$~K for $P=1.8$~GPa and $P=14.7$~GPa. The \textit{a}-lattice was determined from the (220) peak position for all the temperatures and pressures. Together with the measurements of the three-dimensional reciprocal space on the line between the (660) and (664) Bragg peaks and their surroundings, can be concluded that the symmetry and space group of the structure is the same for all studied (\textit{P,T}) combinations, based on the lack of changes in peak shape, peak splitting, as well as the lack of potential additional peaks. 

In Fig.~\ref{Fig1}(d) we present raw x-ray diffraction data of the (664) Bragg peaks for different applied pressures measured at 300~K. The \textit{c}-lattice was determined from the difference between the (664) and (660) peak positions for all the temperatures and pressures. The data are color-coded with the \textit{c}-lattice parameter value [as discussed below for Fig.~\ref{Fig1}(e)]. The foot like feature on the left of the (664) Bragg peak is associated with the crystalline mosaicity which slightly changes for higher pressures.

Figure~\ref{Fig1}(e) shows the temperature-pressure paths  during the experiment. Their order is indicated by the numbers and arrows. At every point, data were collected from which the \textit{a}- and \textit{c}-lattice parameters were inferred. The open symbols indicate the temperatures/pressures at which data were taken and are color-coded with the \textit{c}-lattice parameter value. The pressure cell was gas-loaded at $P=1.1$~GPa which increased to $P=2.8$~GPa while cooling to base temperature (path 1). The sample was then warmed up to 300~K while taking data and cooled back down to 160~K (path 2), at which point the compression membrane was engaged, and the pressure was gradually increased to $P=7$~GPa (path 3). The sample was then cooled down to base temperature (path 4) and warmed to 300~K with pressure staying roughly constant up to 170~K and then reaching $P=14$~GPa upon warming to 300~K (path 5). Finally, at 300~K, the pressure was gradually decreased down to $P=1.7$~GPa, by using the de-compression membrane  (path 6). %From the two isothermal paths at $T=160$ and 300~K, the collapsed tetragonal (cT) transition was clearly observed, and the TET/cT phase boundary was determined, indicated by the red line. 

\begin{figure}[htbp]
	\begin{center}
		\includegraphics[width=86mm, trim=0cm 0cm 0cm 0cm]{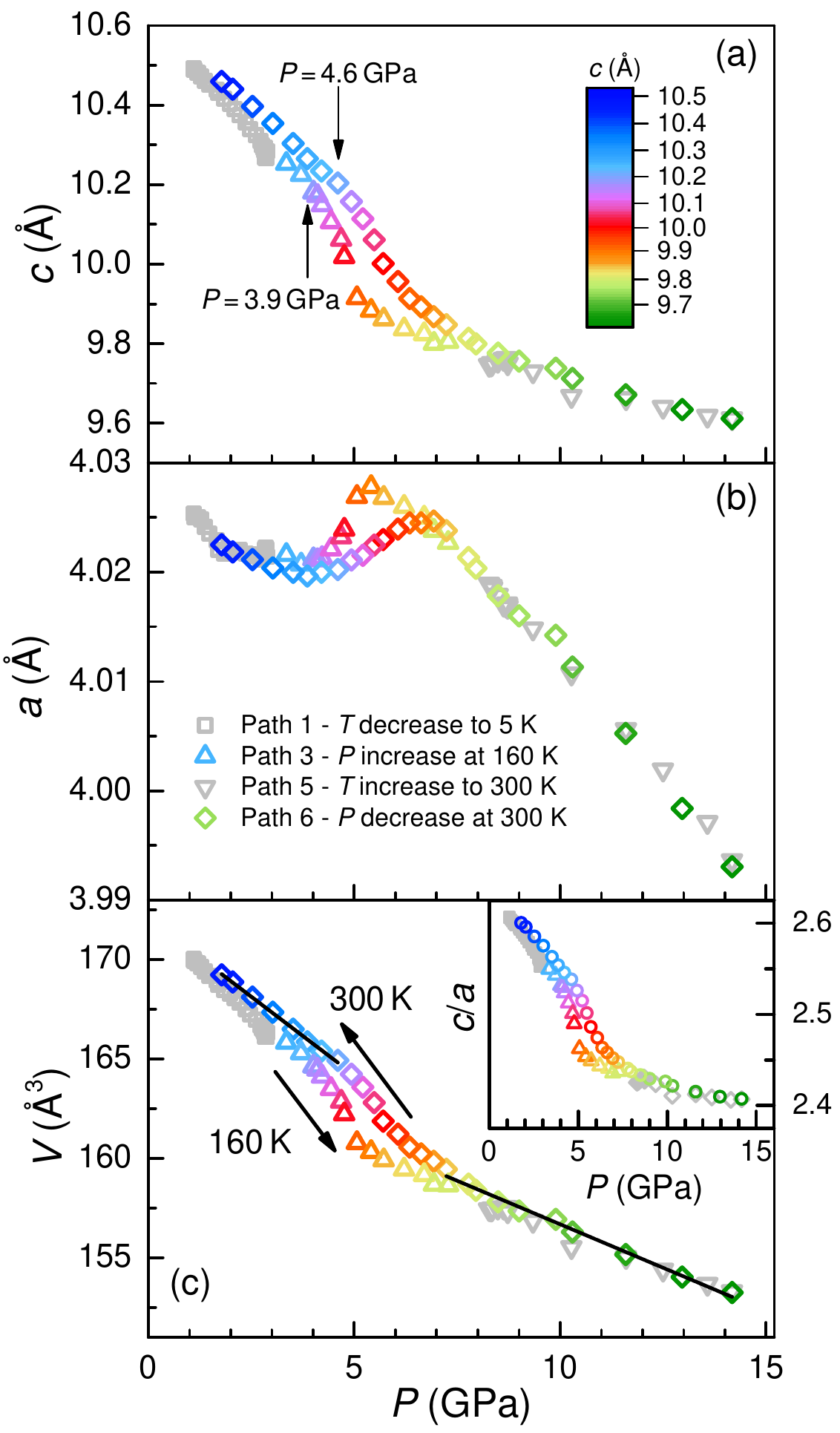}
	\end{center}
	\caption{(a) The inferred \textit{c}-lattice parameter of LaRu$_2$P$_2$ plotted vs. applied pressures (color-coded \textit{c}-lattice parameter). The symbols shape represent the pressure-temperature paths depicted in Fig.~\ref{Fig1}(e). The arrows represent the inferred pressure of the cT transition. The grey symbols represent the non-isothermal paths. (b) The inferred \textit{a}-lattice parameter of LaRu$_2$P$_2$ plotted vs. applied pressure (color-coded \textit{c}-lattice parameter). (c) The inferred unit cell volume of LaRu$_2$P$_2$ plotted vs. applied pressure. Black lines indicate the linear fits used to determine the bulk modulus (\textit{B}) below and above  the cT transition. Inset: \textit{c}/\textit{a} ratio of LaRu$_2$P$_2$ plotted vs. applied pressure.}
	\label{Fig2}
\end{figure}

%In Fig.~\ref{Fig2} we present the \textit{a}- and \textit{c}-lattice parameters. %The data is color-coded with the \textit{c}-lattice parameter value [as discussed below for Fig.~\ref{Fig1}(e)]. 
Figure~\ref{Fig2}(a) shows the \textit{c}-lattice parameter versus applied pressure. During the initial cooling from 300 to 5~K (light grey squares), the sample remained in the TET phase [path 1 in Fig.~\ref{Fig1}(e)]. The \textit{c}-lattice measured in path 1 also allows us to appreciate the relative (larger) change due to pressure compared to thermal contraction. Upon application of pressure at 160~K, the sample is compressed up to the transition in to the cT phase at $P=3.9(3)$~GPa (determined from the kink beyond the linear slope) which is clearly evident in the \textit{c}-lattice parameter size decrease (path 3). With reduction of pressure at 300~K, from $P=14$ to 1.8~GPa, the sample exhibits a transition back to the TET phase at $P=4.6(3)$~GPa (path 5). The arrows represent the pressure at which the cT transition occurs for the two isothermal paths. No change in the symmetry of the (\textit{HK}0) Bragg peaks nor additional peaks were observed when crossing the cT transition [see Fig.~\ref{Fig1}(c)]. %During measurements along path 1, it is possible that the cT phase line was crossed, however we may not be sensitive to it as we might be crossing a "nearly vertical" phase boundary along a vertical line.   

%The diffraction pattern of the (HK0) plane was continuously acquired with the MAR345 image plate detector throughout the experiment.

In Fig.~\ref{Fig2}(b) the \textit{a}-lattice parameter is shown versus applied pressure. Initially, the \textit{a}-lattice parameter decreases until it flattens at the pressure corresponding to the cT transition, peaks and then linearly decreases with increasing pressure beyond the cT transition. A similar pressure dependence for the \textit{a}-lattice parameter was observed in LaFe$_2$P$_2$.\cite{Huhnt1998} Figure~\ref{Fig2}(c) shows the unit cell volume of LaRu$_2$P$_2$ vs. applied pressure, and demonstrates a volume collapse expected for a cT transition.  In the inset of Fig.~\ref{Fig2}(c), the \textit{c}/\textit{a} ratio of LaRu$_2$P$_2$ is shown vs. applied pressure.

\section{Discussion}

The continuous change in the unit cell volume shown in Fig.~\ref{Fig2}(c), suggests that the cT transition is consistent with  a second order transition, similar to the one observed in LaFe$_2$P$_2$, and predicted for LaRu$_2$P$_2$.\cite{Huhnt1998} In addition, for a first-order transition one would expect coexistence of the cT/TET phases which would manifest as splitting or broadening of Bragg peaks~\cite{Huhnt1998,Jayasekara2015, Kaluarachchi2017Pb} which is not evident from the diffraction data in this work [Figs.~\ref{Fig1}(c) and (d)]. Moreover, band-structure calculations predict a continuous change in the lattice parameters~\cite{Li2016}. Thus, we do not expect hysteresis for the cT transition in LaRu$_2$P$_2$ and can define the TET/cT phase boundary between $T=160$ and 300~K, shown in Fig.~\ref{Fig1}(e). The extrapolated TET/cT phase boundary suggests that at $T=0$, one can expect the cT transition to occur near 3~GPa.  In addition, recent measurements of the elastic properties of LaRu$_2$P$_2$ have shown a characteristic behavior consistent with a cT transition.~\cite{Sypek2017}. Moreover, although the sample undergoes an approximately 3\% change in volume at the cT transition, it remains single-crystalline across the volume collapse. This is clearly demonstrated in the raw diffraction data in Fig.~\ref{Fig1}(c), below and above the cT transition, as there are no evidence for the sample breaking into multiple crystallites which would result in the Bragg peaks radially smearing into Bragg rings.

The previously reported pressure dependence of the superconducting properties~\cite{Foroozani2014,Li2016} can be explained by the proximity to the cT phase; namely, the increase and decrease of $T_c$, broadening of the superconducting transition~\cite{Li2016}, and reported loss of the Meissner response above $P=2.1$~GPa\cite{Foroozani2014}. The cT transition extrapolates to 3~GPa at $T=0$. At first glance data measure in path 1 [Fig.~\ref{Fig1}(e))] seems to preclude the cT phase for $P<2.8$~GPa, however the onset criteria may make it hard to detect. Even if $P_{\mathrm{cT}} >2.1$~GPa at 5~K, the loss of SC can be related to the proximity of SC to a structural instability promoted by the cT phase, which initially enhances the electron-phonon coupling and leads to an enhancement of $T_c$ with pressure. Closer to the cT phase, P-P bond fluctuations may alter the electronic density of states~\cite{Hoffmann1985} and ultimately the cT transition will lead to a new band-structure and phonon spectrum that may not support the SC state. Either way, our data are consistent with the disappearance of SC close to, or at the cT phase transition.    

From the volume change with pressure, the isothermal compressibility (and bulk modulus \textit{B}) of LaRu$_2$P$_2$ can be inferred, by linear fitting the unit cell volume shown in Fig.~\ref{Fig2}(c) (black lines), below and above the cT transition. They were found to be $\beta_T = 9.5(2) \times 10^{-3}~\mathrm{GPa}^{-1}$ [$B=105(2)$~GPa] and $\beta_T = 5.7(2) \times 10^{-3}~\mathrm{GPa}^{-1}$ [$B=175(5)$~GPa] for pressures below and above the $P_{CT}$ for the CT transition, respectively. A doubling of the bulk modulus at pressures above the cT transition had been also reported in EuCo$_2$As$_2$.~\cite{Bishop2010}

\begin{figure}[htbp]
	\begin{center}
		\includegraphics[width=86mm]{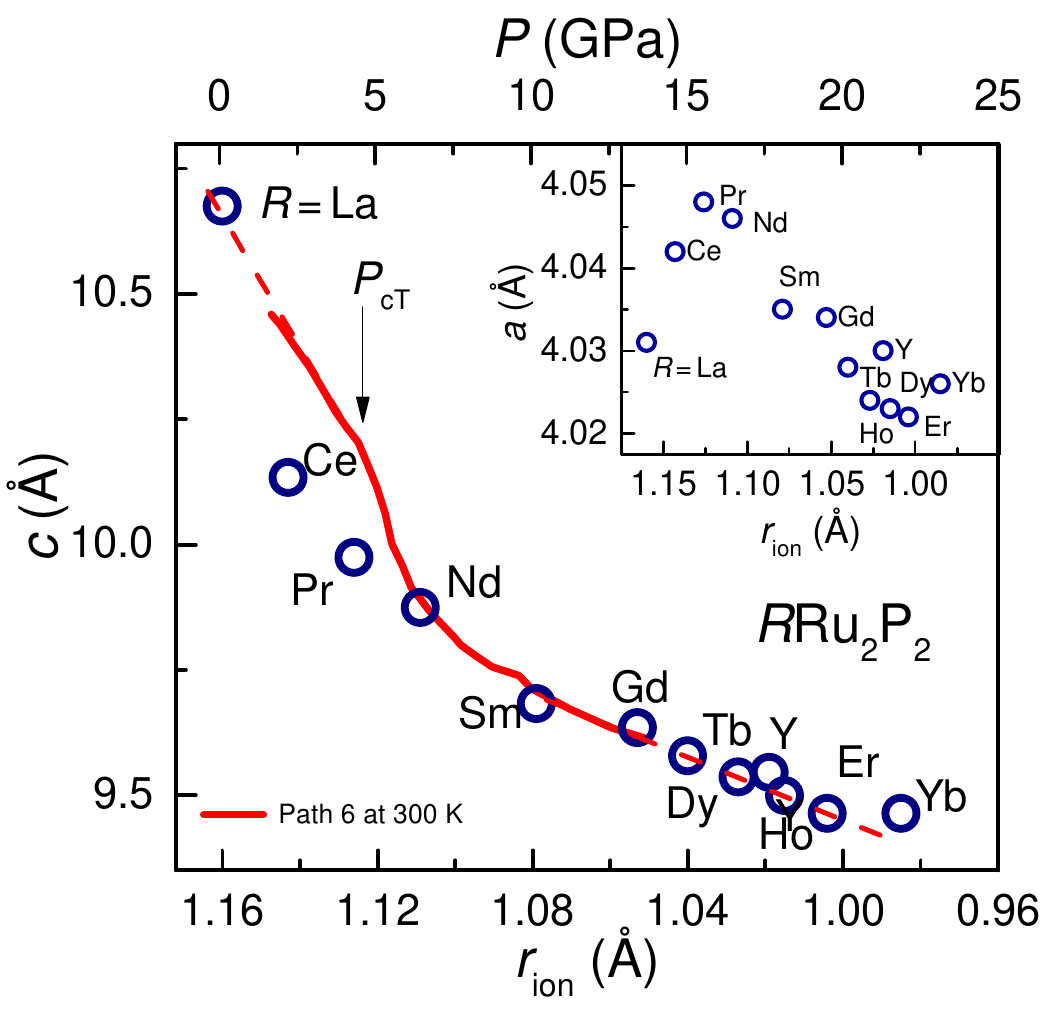}
	\end{center}
	\caption{(a) The \textit{c}-lattice parameter of the \textit{R}Ru$_2$P$_2$ (\textit{R}~=~Y, La$\textendash$Er, Yb) series, for different rare-earth members from Ref.~\cite{Jeitschko1987}, plotted vs. ionic radii, $r _{\mathrm{ion}}$ (bottom \textit{x}-axis), superimposed with LaRu$_2$P$_2$ \textit{c}-lattice parameters measured at $T=300$~K [Path 6 in Fig.~\ref{Fig1}(e)] plotted vs. external pressure \textit{P} (top \textit{x}-axis). The dashed lines are a linear extrapolation to the data. Inset: The \textit{a}-lattice parameter of the \textit{R}Ru$_2$P$_2$ series, for different rare-earth members from Ref.~\cite{Jeitschko1987}, plotted vs. ionic radii.}
	\label{Fig3}
\end{figure}

The pressure dependence of the \textit{c}-lattice parameters of LaRu$_2$P$_2$ at 300~K can be related to the reported size of \textit{c}-lattice parameters of the \textit{R}Ru$_2$P$_2$ (\textit{R} = Y, La$\textendash$Er, Yb) series~\cite{Jeitschko1987} and the effect of external and chemical pressure associated with the lanthanide contraction. Figure~\ref{Fig3} shows the \textit{c}-lattice parameters of the \textit{R}Ru$_2$P$_2$~\cite{Jeitschko1987} as a function of ionic radii ($r _{\mathrm{ion}}$) for a coordination number 8 of the \textit{R} ions~\cite{Shannon1976}. We plot the \textit{c}-lattice parameter measured in path 6 ($T=300$~K) vs \textit{P} on the top axis. The \textit{x}-axes are scaled so that \textit{R} = La is set as $P=0$ and the measured data were linearly extrapolated through the end members of the series. The \textit{c}-lattice parameter change with rare-earth substitution exhibits the same trend as with external pressure applied on LaRu$_2$P$_2$. Moving from \textit{R} = La to \textit{R} = Sm, the unit cell undergoes a cT transition, between \textit{R} = La and Nd. Assuming linear scaling between the chemical pressure associated with the lanthanide contraction and change in ionic radii from \textit{R} = La to \textit{R} = Er, we can estimate that $\frac{{\mathrm{d}{P_{\mathrm{chemical}}}}}{{\mathrm{d}{r_{\mathrm{ion}}}}} \approx  -125$~GPa/$\AA$. We note that scaling between external and chemical pressure works well on both sides and while going through the isostructural cT phase transition.
In the inset of Fig.~\ref{Fig3} we show the \textit{a}-lattice parameters of the \textit{R}Ru$_2$P$_2$ series. Similarly to external pressure, the \textit{a}-lattice parameter vs. $r _{\mathrm{ion}}$ peaks at \textit{R} = Pr, after which it decreases linearly, which strongly supports the analogous effect of chemical and physical pressure in the \textit{R}Ru$_2$P$_2$ series.   

%http://abulafia.mt.ic.ac.uk/shannon/ptable.php
\section{Conclusion}

In summary, we studied the structural properties of LaRu$_2$P$_2$ under applied external pressures up to 14~GPa, using high-energy x-ray diffraction. We find that LaRu$_2$P$_2$ undergoes a cT phase transition, consistent with a second order transition, which is temperature dependent, increasing from $P=3.9(3)$~GPa at 160~K to $P=4.6(3)$~GPa at 300~K. We find that the reported change in the high-pressure superconducting properties in LaRu$_2$P$_2$ is likely driven by the transition to the cT phase. We also determined the bulk modulus of LaRu$_2$P$_2$ to be $B=105(2)$~GPa and $B=175(5)$~GPa, for pressures below and above the $P_{CT}$ for the CT transition, respectively. Finally, we compared the effect of physical and chemical pressure on the lattice parameters of the \textit{R}Ru$_2$P$_2$ (\textit{R} = Y, La$\textendash$Er, Yb) isostructural series of compounds, and found that physical pressure is analogues to the chemical pressure associated with the lanthanide contraction.

\section*{Acknowledgment}

We thank Anna E. B\"{o}hmer for useful and delightful discussions.
The authors would like to acknowledge D.S. Robinson, S. Tkachev, M. Baldini and S.G. Sinogeikin for their assistance. G.D. was funded by the Gordon and Betty Moore Foundation’s EPiQS Initiative through Grant GBMF4411. Work done at Ames Laboratory was supported by US Department of Energy, Basic Energy Sciences, Division of Materials Sciences and Engineering under Contract No. DE-AC02-07CH111358.  This research used resources of the Advanced Photon Source, a U.S. Department of Energy (DOE) Office of Science User Facility operated for the DOE Office of Science by Argonne National Laboratory under Contract No. DE-AC02-06CH11357. We gratefully acknowledge support by HPCAT (Sector 16), Advanced Photon Source (APS), Argonne National Laboratory. HPCAT operations are supported by DOE-NNSA under Award No. DE-NA0001974, with partial instrumentation funding by NSF. Use of the COMPRES-GSECARS gas loading system was supported by COMPRES under NSF Cooperative Agreement EAR 11-57758 and by GSECARS through NSF grant EAR-1128799 and DOE grant DE-FG02-94ER14466. 

\bibliography{Bibliography} 
\end{document}